\documentclass[12pt]{article}
\title{\bf Evaluation of Self-Intersecting Wilson Loop\\
in the Stochastic Vacuum Model}
\author{Dmitri Antonov \thanks{Permanent address:
Institute of Theoretical and Experimental Physics, 
B. Cheremushkinskaya 25, RU-117 218 Moscow, Russia.}{\,}
\thanks{Tel.: + 39 050 844 536; Fax: + 39 050 844 538; 
E-mail address: {\tt antonov@df.unipi.it}} 
\\
{\it INFN-Sezione di Pisa, Universit\'a degli studi di Pisa,}\\
{\it Dipartimento di Fisica, Via Buonarroti, 2 - Ed. B - 
I-56127 Pisa, Italy}}
\date{}
\begin{document}
\maketitle
\vspace{1mm}
\centerline{\bf {Abstract}}
\vspace{3mm}
\noindent
A Wilson loop is evaluated within the 
stochastic vacuum model for the case when the respective contour is 
self-intersecting and its size does not exceed 
the correlation length of the vacuum. 
The result has the form of a certain functional of the tensor area. 
It is similar to that for the non-self-intersecting loop only 
when the contour is a plane one. Even for such a contour, 
the obtained expression depends on the ratio of two
functions parametrizing the bilocal field strength
correlator taken at the origin, which is not so for the case of 
non-self-intersecting contour. 

\vspace{3mm}

\vspace{3mm}

\newpage

The small non-self-intersecting 
Wilson loop in QCD~\footnote{From now on, 
we shall for brevity call by self-intersecting loop the loop defined at 
a self-intersecting contour and imply by the size of the loop the size of 
this contour. Besides that, all the investigations will be performed 
in the Euclidean space-time.}, 

$$\left<W(C)\right>\equiv\frac{1}{N_c}\left<{\rm tr}{\,}{\cal P}{\,}\exp
\left(ig\oint dx_\mu A_\mu^aT^a\right)\right>,$$
has been 
evaluated long time ago. First, it followed indirectly from the 
respective quark-antiquark potential emerging due to the dipole interaction 
with the external colourelectric field~\cite{1} and then was found 
explicitly by performing the direct calculation in QCD~\cite{3}. 
The term ``small'' here means that
the typical size of the loop does
not exceed the correlation length of QCD vacuum, $T_g$, {\it i.e.} the 
distance at which the bilocal field strength correlator 
in stochastic vacuum model~\cite{3,4} decreases. This length has been 
measured in the lattice experiments~\cite{5,6} (see Refs.~\cite{7,8}
for reviews) with the result $T_g\simeq 0.34{\,}{\rm fm}$ for the 
realistic case of
$SU(3)$ full QCD. As a consequence of smallness 
of the loop 
{\it w.r.t.} $T_g$, the field strength correlator 
of stochastic vacuum model can with a 
good accuracy be approximated during the evaluation of such a loop 
by gluonic condensate. Therefore
for small loops the QCD vacuum is viewed as that of 
QCD sum rules~\cite{9} characterized by the infinite correlation length. 
However, self-intersecting loops are also of a great importance for 
QCD, since these are those loops at which loop equations~\cite{10} are 
nontrivial. In particular, in 2D QCD such loops have been 
comprehensively studied in Ref.~\cite{11} (see Ref.~\cite{lect} for a 
review).
In the present letter, by combining stochastic vacuum model
with the loop space approach we shall evaluate small 
self-intersecting Wilson loop in 4D QCD with arbitrary number 
of colours. As we shall eventually see, the resulting expression 
differs significantly from that for a non-self-intersecting loop.

The idea we are going to employ 
is based on the possibility to represent the loop-space 
Laplacian~\cite{12} 

\begin{equation}
\label{1}
\Delta\equiv\int\limits_{0}^{1}d\sigma\int
\limits_{\sigma-0}^{\sigma+0}d\sigma'\frac{\delta^2}{\delta x_\mu(\sigma')
\delta x_\mu(\sigma)}
\end{equation} 
standing on the L.H.S. of the 
loop equations in the following form~\cite{13}:

\begin{equation}
\label{2}
\Delta=\oint dx_\mu(\sigma){\,} \textsf{v.p.}\int d\sigma'
\dot x_\nu(\sigma')
\frac{\delta^2}{\delta\sigma_{\lambda\mu}(x(\sigma))
\delta\sigma_{\nu\lambda}(x(\sigma'))},
\end{equation}
where $\textsf{v.p.}\int d\sigma'\equiv
\int\limits_{0}^{\sigma-0}d\sigma'+
\int\limits_{\sigma+0}^{1}d\sigma'$. 
It is worth mentioning once more that as it follows from the
loop equations, the result of the action of the loop-space Laplacian  
onto the Wilson loop is nonvanishing only provided that 
this loop is self-intersecting. By virtue of Eq.~(\ref{2}), we get for
such a loop the following equation:

\begin{equation}
\label{4}
\Delta\left<W(C)\right>=-\frac{g^2}{N_c}
\oint dx_\mu(\sigma){\,} \textsf{v.p.}\int d\sigma'
\dot x_\nu(\sigma'){\,}{\rm tr}{\,}\left<F_{\lambda\mu}(x)
\Phi(x,x')F_{\nu\lambda}(x')\Phi(x',x)\right>.
\end{equation}
Here, $\Delta$ is defined by Eq.~(\ref{1}), $x\equiv x(\sigma)$, 
$x'\equiv x(\sigma')$, 
$F_{\mu\nu}=\partial_\mu A_\nu-\partial_\nu A_\mu-ig[A_\mu, A_\nu]$
stands for the Yang-Mills field strength tensor, with 
$A_\mu\equiv A_\mu^aT^a$,
and $\Phi(x,y)\equiv
\frac{1}{N_c}{\cal P}\exp\left(ig\int\limits_{y}^{x}
A_\mu(u)du_\mu\right)$ is a parallel transporter factor along the 
respective part of the contour $C$. We can now take into account the fact
that the characteristic size of the countour $C$ under consideration 
does not exceed $T_g$~\footnote{More rigorously, this means that 
the area of the minimal surface spanned by $C$ is not larger than 
$T_g^2$ and that $C$ does not contain appendix-shaped pieces of 
the vanishing area, but large length.}. This enables us to use for
the field strength correlator standing on the R.H.S. of Eq.~(\ref{4})
the expression known from the stochastic vacuum model~\cite{3,4}.
In fact, for such a small contour joining the points $x$ and $x'$
the result for the bilocal field strength correlator suggested 
by this model is independent of the form of the contour and reads

$$
\frac{g^2}{2}\left<F_{\mu\nu}(x)\Phi(x,x')
F_{\lambda\rho}(x')\Phi(x',x)\right>=\frac{\hat 1_{N_c\times 
N_c}}{N_c}\left\{(\delta_{\mu\lambda}\delta_{\nu\rho}-\delta_{\mu\rho}
\delta_{\nu\lambda}){\cal D}(x-x')+\right.$$

\begin{equation}
\label{dd1}
\left.+\frac12\left[\partial_\mu^x((x-x')_\lambda
\delta_{\nu\rho}-(x-x')_\rho\delta_{\nu\lambda})+
\partial_\nu^x((x-x')_\rho\delta_{\mu\lambda}-(x-x')_\lambda
\delta_{\mu\rho})\right]{\cal D}_1(x-x')\right\}.
\end{equation}

In what follows, we shall consider only nonperturbative 
parts of the functions ${\cal D}$ and ${\cal D}_1$, which were measured 
on the lattice in Refs.~\cite{5,6,7,8}.
That is because the  
perturbative contributions to these functions, 
relevant to the UV divergencies~\cite{markus}, yield 
a renormalization factor, which cancels out 
with the same factor appearing on the L.H.S. of Eq.~(\ref{4}) during 
the direct renormalization of the 
Wilson loop~\footnote{The multiplicative 
renormalizability of self-intersecting Wilson loop has been proved 
in Ref.~\cite{brandt}.}. 
Therefore, from now on 
we shall deal with the renormalized Wilson loop as well as the 
renormalized charge $g$. Owing to Eq.~(\ref{dd1}), Eq.~(\ref{4})
takes the form $\Delta W[x]=J[x]$, where $W[x]\equiv \left<W(C)\right>$
and $J[x]=\oint dx_\mu(\sigma){\,} \textsf{v.p.}\int d\sigma'
\dot x_\mu(\sigma')F(x-x')$ with 

\begin{equation}
\label{Func}
F(x)\equiv\frac{1}{N_c}\left[6{\cal D}(x)+
4{\cal D}_1(x)+x_\mu\partial_\mu{\cal D}_1(x)\right].
\end{equation}

The above equation for the Wilson loop can 
be solved by virtue of the method of inversion of the loop-space 
Laplacian proposed in Ref.~\cite{lapl}~\footnote{This method has been 
successfully applied in Ref.~\cite{turb} 
to the solution of the Cauchy problem for the 
loop equation in turbulence.}. The idea of this method is to replace the 
original loop-space Laplacian~(\ref{1}) by the smeared one 
$\Delta^{(G)}=\int\limits_{0}^{1}
d\sigma\int\limits_{0}^{1}d\sigma'G(\sigma-\sigma')\frac{\delta^2}{
\delta x_\mu(\sigma')\delta x_\mu(\sigma)}$, where 
$G(\sigma-\sigma')$ is a certain smearing function. Such a smeared Laplacian 
can be inverted, which yields

\begin{equation}
\label{res}
W[x]=1-\frac12\int
\limits_{0}^{\infty}dA\left(\left<J\left[x+\sqrt{A}
\xi\right]\right>_\xi^{(G)}-
\left<J\left[\sqrt{A}\xi\right]\right>_\xi^{(G)}\right).
\end{equation}
Here, 

$$\left<{\cal O}[\xi]\right>_\xi^{(G)}=\frac{
\int\limits_{\xi(0)=\xi(1)}^{}
{\cal D}\xi{\rm e}^{-S}{\cal O}[\xi]}{\int\limits_{\xi(0)=\xi(1)}^{}
{\cal D}\xi{\rm e}^{-S}}$$ 
is the Gaussian average over loops with 
the action 
$S=\frac12\int\limits_{0}^{1}d\sigma\int\limits_{0}^{1}d\sigma'\xi(\sigma)
G^{-1}(\sigma-\sigma')\xi(\sigma')$, where 
$G^{-1}(\sigma-\sigma')$ denotes the inverse operator.
Following Ref.~\cite{lapl}, we shall choose $G(\sigma-\sigma')$ in the form 
$G(\sigma-\sigma')={\rm e}^{-|\sigma-\sigma'|/\varepsilon}$, 
$\varepsilon\ll 1$,
after which the above action becomes local: $S=\frac14
\int\limits_{0}^{1}d\sigma\left(\varepsilon\dot\xi^2(\sigma)+
\frac{1}{\varepsilon}\xi^2(\sigma)\right)$. 

Next, the averages on the 
R.H.S. of Eq.~(\ref{res}) are similar to those which one carries out
during the reproduction of the one-gluon-exchange diagram within the 
same method when it is applied to the usual loop equation in the 
large-$N_c$ limit~\cite{lapl}. In particular, this follows from the 
fact that the integral operator present in $J[x]$ accidentally has 
the same form as the one
standing on the R.H.S. of the large-$N_c$ loop equation.
In the limit $\varepsilon\to 0$, the only nonvanishing
contribution appears in the first of the two averages on the 
R.H.S. of Eq.~(\ref{res}) and reads

$$W[x]=1-\frac12\int\limits_{0}^{\infty}dA
\int\limits_{0}^{1}
d\sigma\int\limits_{0}^{1}d\sigma'(1-G(\sigma-\sigma'))\dot x_\mu(\sigma)
\dot x_\mu(\sigma')\times$$

\begin{equation}
\label{inter}
\times\int\frac{d^4p}{(2\pi)^4}
\exp\left[-Ap^2(1-G(\sigma-\sigma'))+ip(x(\sigma)-x(\sigma'))\right]
\tilde F(p).
\end{equation}
Here, the factor 
${\rm e}^{-Ap^2(1-G(\sigma-\sigma'))}$ is the result of the average
$\left<{\rm e}^{i\sqrt{A}p(\xi(\sigma)-\xi(\sigma'))}\right>_\xi^{(G)}$, 
and $\tilde F(p)\equiv\int d^4x{\rm e}^{-ipx}F(x)$ is the Fourier image 
of the function $F(x)$. Next, in Eq.~(\ref{inter}), 
the factor $(1-G(\sigma-\sigma'))$ 
in the preexponent was introduced in order to make out of the full 
integral over $\sigma'$ the principal-value one in the limit
$\varepsilon\to 0$. This factor disappears upon the introduction instead
of $A$ the new integration variable $\alpha=\Lambda^{-2}+2A
(1-G(\sigma-\sigma'))$, where $\Lambda$ stands for the UV momentum 
cutoff. Sending $\Lambda$ to infinity we arrive at the following expression:

$$W[x]=1-\frac12\oint dx_\mu\oint dx_\mu'\int\frac{d^4p}{(2\pi)^4}
\frac{{\rm e}^{ip(x-x')}}{p^2}\tilde F(p)=$$

\begin{equation}
\label{rez}
=1-\frac{1}{8\pi^2}\oint dx_\mu\oint dx_\mu'\int d^4z
\frac{F(z)}{(z-(x-x'))^2}.
\end{equation}

The infinite integral ({\it i.e.} the integral over $z$ or $p$) 
can be calculated in the 
small-loop case under study. In this limit, one can replace ${\cal D}(x)$ and 
${\cal D}_1(x)$ by their values at the origin, which according 
to Eq.~(\ref{dd1}) are related to each other as 

\begin{equation}
\label{rel}
{\cal D}(0)+{\cal D}_1(0)=
\frac{g^2}{24}{\,}{\rm tr}{\,}\left<F_{\mu\nu}^2(0)\right>.
\end{equation} 
On the 
other hand, according to the lattice measurements~\cite{5,6,7,8}, 
${\cal D}_1(0)=\alpha{\cal D}(0)$, where $\alpha\simeq 0.2\pm 0.1$ 
(see also Ref.~\cite{wint} for the discussion of this value of $\alpha$).
This yields the following approximate constant value of the function 
$F(z)$: $F(z)\simeq 
\frac{3+2\alpha}{1+\alpha}\frac{g^2}{12N_c}
{\,}{\rm tr}{\,}\left<F_{\mu\nu}^2(0)\right>\equiv{\cal C}$. 
The remaining infinite 
integral can easily be calculated in the limit $T_g\gg |x-x'|$
under study {\it e.g.} from the first equality on the R.H.S. of 
Eq.~(\ref{rez}). We have 

$$\int\frac{d^4p}{(2\pi)^4}
\frac{{\rm e}^{ip\lambda}}{p^2}\tilde F(p)={\cal C}
\int d^4p\delta(p)\frac{{\rm e}^{ip\lambda}}{p^2}
\simeq\frac{{\cal C}T_g^4}{16\pi^2}\int\limits_{0}^{\infty}ds\int d^4p
{\rm e}^{ip\lambda-p^2\left(\frac{T_g^2}{4}+s\right)}=
\frac{{\cal C}T_g^4}{4\lambda^2}
\left(1-{\rm e}^{-\frac{\lambda^2}{T_g^2}}\right),$$
where $\lambda\equiv x-x'$, and thus 

$$
W[x]=1-\frac{{\cal C}T_g^4}{8}
\oint dx_\mu\oint dx_\mu'\frac{1}{(x-x')^2}
\left[1-{\rm e}^{-\frac{(x-x')^2}{T_g^2}}\right].$$
Expanding the exponential in this formula we finally arrive at the 
following leading nontrivial contribution to the Wilson loop:

\begin{equation}
\label{lead}
\left<W(C)\right>\simeq 1-\frac{3+2\alpha}{1+\alpha}\frac{g^2}{96N_c}
{\,}{\rm tr}{\,}\left<F_{\mu\nu}^2(0)\right>
\Sigma_{\mu\nu}^2,
\end{equation} 
where 
$\Sigma_{\mu\nu}\equiv\oint dx_\mu x_\nu$ is the tensor area 
corresponding to the contour $C$. Note that $T_g$ dropped out from this 
leading term, as it could be expected from the beginning, since 
$1/T_g$ was considered as an IR cutoff. 

The obtained expression for the Wilson loop is now worth to be 
compared with the respective expression for the small 
non-self-intersecting Wilson loop. In that case 
owing to the non-Abelian Stokes theorem and Eq.~(\ref{dd1}) one has

$$\left<W(C)\right>\simeq\frac{1}{N_c}{\,}{\rm tr}{\,}\exp\left(
-\frac{g^2}{8}\int\limits_{\Sigma_{\rm min}[C]}^{}d\sigma_{\mu\nu}(x)
\int\limits_{\Sigma_{\rm min}[C]}^{}d\sigma_{\lambda\rho}(x')
\left<F_{\mu\nu}(x)\Phi(x,x')
F_{\lambda\rho}(x')\Phi(x',x)\right>\right)=$$

$$=\exp\left(-\frac{g^2}{48N_c}
{\,}{\rm tr}{\,}\left<F_{\mu\nu}^2(0)\right>
\int\limits_{\Sigma_{\rm min}[C]}^{}d\sigma_{\mu\nu}(x)
\int\limits_{\Sigma_{\rm min}[C]}^{}d\sigma_{\mu\nu}(x')\right).$$
Here, $x\equiv x(\xi)$ is the vector parametrizing the 
surface of the minimal area spanned by the contour $C$, 
$\Sigma_{\rm min}[C]$, and  
$\xi=\left(\xi^1,\xi^2\right)$ stands for the 2D-coordinate. 
Next, $d\sigma_{\mu\nu}(x)=\sqrt{g(\xi)}
t_{\mu\nu}(\xi)d^2\xi$, where $g(\xi)$ is the determinant of the 
induced metric tensor $g^{ab}(\xi)=(\partial^a x_\mu(\xi))
(\partial^b x_\mu(\xi))$ and $t_{\mu\nu}(\xi)=\varepsilon^{ab}
\left(\partial_a x_\mu(\xi)\right)
\left(\partial_b x_\nu(\xi)\right)/\sqrt{g(\xi)}$ is the extrinsic 
curvature tensor. Since the contour $C$ under discussion is very small
(and consequently the same is $\Sigma_{\rm min}[C]$), 
the points $x$ and $x'$ are located very closely to each other, and 
therefore $t_{\mu\nu}(\xi)t_{\mu\nu}(\xi')\simeq t_{\mu\nu}^2(\xi)=2$.
Finally, taking into account that $\int d^2\xi\sqrt{g(\xi)}=
{\,}{\rm Area}{\,}{\rm of}{\,}\Sigma_{\rm min}[C]\equiv S_{\rm min}$, 
we obtain~\cite{3} $\left<W(C)\right>\simeq 1-
\frac{g^2}{24N_c}
{\,}{\rm tr}{\,}\left<F_{\mu\nu}^2(0)\right>S_{\rm min}^2$. 
For a {\it plane} contour, this 
expression has the form similar to our Eq.~(\ref{lead}), since for 
such a contour 

\begin{equation}
\label{ar}
S_{\rm min}^2=\frac12\Sigma_{\mu\nu}^2.
\end{equation} 
In this case, 
the main difference between these two expressions stems from the 
$\alpha$-dependence of Eq.~(\ref{lead}). This dependence is due to the 
fact that the functions ${\cal D}$ and ${\cal D}_1$  
contribute to the self-intersecting loop in the nontrivial 
combination~(\ref{Func}), whereas the non-self-intersecting loop 
depends only on their sum at the origin, expressible 
owing to Eq.~(\ref{rel}) via the 
condensate alone. Another obvious difference 
of the two expressions for Wilson loops is that the tensor area
for self-intersecting countour can be vanishingly small even 
for a very large contour (although we do not consider such contours)
and even vanish completely for the eight-shaped contour with 
equal petals, whereas for a non-self-intersecting contour  
it could vanish only together with the contour itself. Moreover,
in this respect 
it is worth pointing out once more that the comparison of the 
results for self-intersecting 
and non-self-intersecting Wilson loops is only possible for plane 
contours, since  
for non-plane ones Eq.~(\ref{ar}) is not valid. 

In conclusion, by making use of the method of inversion of the 
loop-space Laplacian, 
we have restored a small self-intersecting Wilson loop 
from the bilocal field strength correlator of 
stochastic vacuum model. There turned out to be two main  
differences of the obtained result~(\ref{lead}) from that of 
the non-self-intersecting loop. Firstly, Eq.~(\ref{lead}) 
depends on the tensor area
of the contour, rather than on the area of the minimal surface,
and can therefore vanish for some class of contours ({\it e.g.} for
plane eight-shaped contours with 
equal petals). Secondly, the obtained result 
depends on the ratio of nonperturbative parts of the 
functions ${\cal D}_1(x)$ and ${\cal D}(x)$ at the origin, which is 
not the case for a non-self-intersecting contour. However, 
for plane contours, the 
functional form of the obtained result coincides 
with that of a small non-self-intersecting loop when the latter one is 
expressed in terms of the tensor area.

\section*{Acknowledgments}
The author is indebted to Profs.  
A. Di Giacomo, H.G. Dosch, and Yu.M. Makeenko for 
useful discussions. He is also greatful to Prof. A. Di Giacomo and 
the whole staff of the Quantum Field Theory Division
of the University of Pisa for cordial hospitality and to INFN for the  
financial support.

\end{document}